\documentstyle[times,balanced,epsf,prl,aps]{revtex}

\begin{document}

\draft

\title{Dynamics of a Small Neutrally Buoyant Sphere in a Fluid and 
Targeting in Hamiltonian Systems}

\author{
Armando Babiano$^{1,}$\cite{bemail}
,
Julyan H. E. Cartwright$^{2,}$\cite{jemail}
,  
Oreste Piro$^{3,}$\cite{oemail}
, and
Antonello Provenzale$^{4,}$\cite{aemail}
} 

\address{
$^2$ Laboratoire de M\'et\'eorologie Dynamique, \'Ecole Normale Sup\'erieure,
F-75231 Paris, France \\
$^2$Laboratorio de Estudios Cristalogr\'aficos, IACT (CSIC-UGR), 
E-18071 Granada, Spain \\
$^3$Institut Mediterrani d'Estudis Avan\c{c}ats, IMEDEA (CSIC--UIB),
E-07071 Palma de Mallorca, Spain \\ 
$^4$Istituto di Cosmogeofisica, CNR, I-10133 Torino, Italy
}

\date{Physical Review Letters, {\bf 84}, 5764--5767, 2000}

\maketitle

\begin{abstract}
We show that, even in the most favorable case, the motion of a small spherical 
tracer suspended in a fluid of the same density may differ from the
corresponding motion of an ideal passive particle. We demonstrate furthermore
how its dynamics may be applied to target trajectories in Hamiltonian systems.
\end{abstract}


\begin{twocolumns}

We show with the simplest model for the force acting on a small rigid neutrally
buoyant spherical tracer particle in an incompressible two-dimensional fluid
flow that tracer trajectories separate from fluid trajectories in those regions
where the flow has hyperbolic stagnation points. A tracer will only evolve on
fluid trajectories with Lyapunov exponents bounded by the value of its
reciprocal Stokes number. By making the Stokes number large enough, one can
force a tracer in a flow with chaotic pathlines to settle on either the regular
KAM-tori dominated regions or to selectively visit the chaotic regions with
small Lyapunov exponents. These findings should be of interest for the
interpretation of Lagrangian observations, for example in oceanography, and in
laboratory fluid experiments that use small neutrally buoyant tracers.
Moreover, since a two-dimensional incompressible flow is a particular instance
of a generically chaotic Hamiltonian system, our results constitute a tool for
targeting trajectories in Hamiltonian systems.

Our starting point is the equation of motion for a small rigid 
spherical tracer in an incompressible fluid,
\begin{eqnarray}
&\rho_p\displaystyle\frac{d{\mathbf v}}{dt}=&
\rho_f\displaystyle\frac{D{\mathbf u}}{Dt}+(\rho_p-\rho_f){\mathbf g}
\nonumber \\
&&-\displaystyle\frac{9\nu\rho_f}{2a^2}
\left({\mathbf v}-{\mathbf u}
-\displaystyle\frac{a^2}{6}\nabla^2{\mathbf u}\right) \nonumber \\
&&-\displaystyle\frac{\rho_f}{2}\left(\displaystyle\frac{d{\mathbf v}}{dt}-
\displaystyle\frac{D}{Dt}\left[{\mathbf u}
+\displaystyle\frac{a^2}{10}\nabla^2{\mathbf u}\right]\right) 
\nonumber \\
&&-\displaystyle\frac{9\rho_f}{2a}\displaystyle\sqrt\frac{\nu}{\pi}
\displaystyle\int_{0}^{t}
\displaystyle\frac{1}{\sqrt{t-\zeta}}\displaystyle\frac{d}{d\zeta}
\left({\mathbf v}-{\mathbf u}
-\displaystyle\frac{a^2}{6}\nabla^2{\mathbf u}\right)d\zeta, \nonumber \\
&&
\label{eom}\end{eqnarray}
where ${\mathbf v}$ represents the velocity of the particle, ${\mathbf u}$ that
of the fluid, $\rho_p$ the density of the particle, $\rho_f$, the density of 
the fluid it displaces, $\nu$, the kinematic viscosity of the fluid, $a$, the
radius of  the particle, and $\mathbf g$, gravity. The terms on the right 
represent respectively the force exerted by the undisturbed flow
on the particle, buoyancy, Stokes drag, the added mass, 
and the Basset--Boussinesq history force \cite{boussinesq,basset}. 
The terms in $a^2\nabla^2{\mathbf u}$ are the Fax\'en corrections
\cite{faxen}. The equation is as given by Maxey \&
Riley \cite{maxey}, except for the added mass term, whose correct form 
was first derived by Taylor \cite{taylor}, as was pointed out by Auton 
et al.\ \cite{auton}. The derivative $D{\mathbf u}/Dt$ is along the path 
of a fluid element
\begin{equation}
\frac{D{\mathbf u}}{Dt}=\frac{\partial{\mathbf u}}{\partial t}
+({\mathbf u}\cdot\nabla){\mathbf u}
,\label{euler}\end{equation}
whereas the derivative $d{\mathbf u}/dt$ 
is taken along the trajectory of the particle
\begin{equation}
\frac{d{\mathbf u}}{dt}=\frac{\partial{\mathbf u}}{\partial t}
+({\mathbf v}\cdot\nabla){\mathbf u}
.\label{lagrange}\end{equation}
Equation (\ref{eom}) is valid where the particle 
radius and its Reynolds number are small, as are the velocity gradients
around it, and the initial velocities of the particle and the fluid are equal. 
An excellent review of the history and physics of this problem
is provided by Michaelides \cite{michaelides}.

We shall simplify Eq. (\ref{eom}) even further with the aim of discovering
whether in the most favorable case a tracer particle may always be faithful to
a flow trajectory. With this in mind, we set $\rho_p=\rho_f$, so that the 
tracer be neutrally buoyant. At the same time we assume that it be sufficiently
small that the Fax\'en corrections be negligible. Furthermore we exclude the
Basset--Boussinesq term, since our approach --- in which we follow Einstein and 
others \cite{einstein,russel,crisanti} --- is to obtain a
minimal model with which we may perform a mathematical analysis of the
problem. If in this model there appear differences between particle
and flow trajectories, with the inclusion of further terms these discrepancies 
will remain or may even be enhanced \cite{druzhinin,tanga,yannacopoulos}.
If we rescale space, time, and velocity by scale factors $L$, $T=L/U$, and $U$, 
we arrive at the expression
\begin{equation}
\frac{d{\mathbf v}}{dt}=\frac{D{\mathbf u}}{Dt}
-{\mathrm St}^{-1}\left({\mathbf v}-{\mathbf u}\right) 
-\frac{1}{2}\left(\frac{d{\mathbf v}}{dt}-\frac{D{\mathbf u}}{Dt}\right)
,\label{neutral}\end{equation}
where ${\mathrm St}$ is the particle Stokes number
${\mathrm St}=2a^2 U/(9\nu L)=2/9\,(a/L)^2 {\mathrm Re}_f$, 
${\mathrm Re}_f$ being
the fluid Reynolds number. The assumptions involved in deriving 
Eq.\ (\ref{eom}) require that ${\mathrm St}\ll 1$. 

In the past it has been assumed that sufficiently small neutrally buoyant 
particles have trivial dynamics \cite{crisanti,druzhinin}, and the mathematical 
argument used to back this up is that if we make the approximation 
$D{\mathbf u}/Dt=d{\mathbf u}/dt$, the problem becomes very simple:
\begin{equation}
\frac{d}{dt}({\mathbf v}-{\mathbf u})=
-\frac{2}{3}\,{\mathrm St}^{-1}({\mathbf v}-{\mathbf u}) 
.\end{equation}
Thence 
$
{\mathbf v}-{\mathbf u}=\left({\mathbf v}_{0}
-{\mathbf u}_{0}\right)\exp(-2/3\,{\mathrm St}^{-1}\,t) 
$,
from which we infer that even if we release the particle with a
different initial velocity ${\mathbf v}_{0}$ to that of the fluid 
${\mathbf u}_{0}$, after a transient phase the
particle velocity will match the fluid velocity, $\mathbf v=\mathbf u$, 
meaning that following this argument, a neutrally buoyant particle should be
an ideal tracer.
Although from the foregoing it would seem that neutrally buoyant particles 
represent a trivial limit to Eq.\ (\ref{eom}), this would be without taking 
into account that in a correct approach to the problem $D{\mathbf u}/Dt\neq
d{\mathbf u}/dt$. If we substitute the expressions for these derivatives given
in Eqs.\ (\ref{euler}) and (\ref{lagrange}) into Eq.\ (\ref{neutral}), we 
obtain
\begin{equation}
\frac{d}{dt}\left({\mathbf v}-{\mathbf u}\right)=
-\left(\left({\mathbf v}-{\mathbf u}\right)\cdot\nabla\right){\mathbf u}
-\frac{2}{3}\,{\mathrm St}^{-1}\left({\mathbf v}-{\mathbf u}\right) 
.\end{equation}
We may then write ${\mathbf A}={\mathbf v}-{\mathbf u}$, whence 
\begin{equation}
\frac{d{\mathbf A}}{dt}=-\left(J+\frac{2}{3}\,{\mathrm St}^{-1} I\right)
\cdot{\mathbf A}
,\label{Aeqn}\end{equation}
where $J$ is the Jacobian matrix --- we shall now concentrate on
two-dimensional flows ${\mathbf u}=(u_x, u_y)$ ---
\begin{equation}
J=
\left(
\begin{array}{cc}
\partial_x u_x & \partial_y u_x \\
\partial_x u_y & \partial_y u_y
\end{array}
\right)
.\end{equation}
If we diagonalize the matrix we obtain
\begin{equation}
\frac{d{\mathbf A}_D}{dt}=
\left(
\begin{array}{cc}
\lambda-2/3\,{\mathrm St}^{-1} & 0 \\
0 & -\lambda-2/3\,{\mathrm St}^{-1}
\end{array}
\right)
\cdot{\mathbf A}_D
,\label{AD}\end{equation}
so if ${\mbox{\it Re}}(\lambda)>2/3\,{\mathrm St}^{-1}$, ${\mathbf A}_D$ may 
grow exponentially. Now $\lambda$ satisfies $\mathrm{det}(J-\lambda I)=0$, so 
$\lambda^2-{\mathrm tr}J+{\mathrm det}J=0$. Since the flow is incompressible,
$\partial_x u_x+\partial_y u_y=\mathrm{tr}J=0$,
thence $-\lambda^2=\mathrm{det}J$. Given squared vorticity  
$\omega^2=(\partial_x u_y-\partial_y u_x)^2$, 
and squared strain $s^2=s_1^2+s_2^2$, where the normal component is
$s_1=\partial_x u_x-\partial_y u_y$ and the shear component is
$s_2=\partial_y u_x+\partial_x u_y$, we may write 
\begin{equation}
Q=\lambda^2=-{\mathrm det}J=(s^2-\omega^2)/4
,\end{equation}
where $Q$ is the Okubo--Weiss parameter \cite{okubo,weiss2}.
If $Q>0$, $\lambda^2>0$, and $\lambda$ is real, so deformation dominates, as
around hyperbolic points, whereas if $Q<0$, $\lambda^2<0$, and $\lambda$ is
imaginary, so rotation dominates, as near elliptic points. Equation (\ref{Aeqn})
together with $d{\mathbf x}/dt={\mathbf A}+{\mathbf u}$ defines a dissipative
dynamical system 
\begin{equation} 
d\mbox{\boldmath$\xi$}/dt={\mathbf F}(\mbox{\boldmath$\xi$}) 
\label{disssys}
\end{equation} 
with constant divergence $\nabla\cdot{\mathbf F}=-4/3\,{\mathrm St}^{-1}$ in 
the four dimensional phase space $\mbox{\boldmath$\xi$}=(x,y,A_x,A_y)$, so that
while small values of ${\mathrm St}$ allow for large values of the divergence, 
large values of ${\mathrm St}$ force the divergence to be small. The Stokes
number is the relaxation time of the particle back onto the fluid trajectories 
compared to the time scale of the flow --- with larger ${\mathrm St}$, the 
particle has more independence from the fluid flow. From Eq.\ (\ref{AD}), about 
areas of the flow near to hyperbolic stagnation points with 
$Q>4/9\,{\mathrm St}^{-2}$, particle and flow trajectories separate 
exponentially.

\begin{figure}
\begin{center}
\def\epsfsize#1#2{0.48\textwidth}
\leavevmode
\epsffile{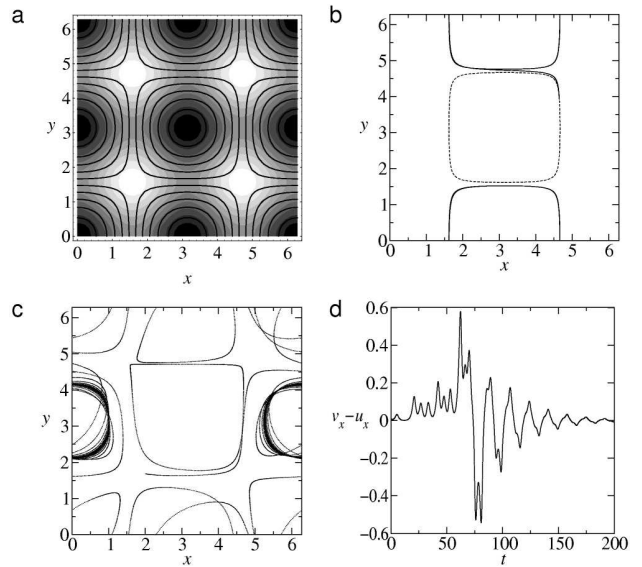}
\end{center}
\caption{\label{timeind}
(a) Fluid trajectories (thick lines) and magnitude of $Q$ 
(background shading: lighter is higher $Q$) 
for the time-independent model Eq.\ (\ref{flow}) with $A=100$ and $B=0$.
The flow is on a torus.
(b) Separation of a neutrally buoyant particle trajectory (solid line) with 
Stokes number ${\mathrm St}=0.2$ from the flow (dashed line) in regions of high 
$Q$ allows the particle to wander between cells.
(c) After a complicated excursion, a particle eventually settles in a zone of 
low $Q$.
(d) The velocity difference $v_x - u_x$ between the particle and the flow 
against time for case (c).
}
\end{figure}

To illustrate the effects of ${\mathrm St}$ and $Q$ on the dynamics of a 
neutrally buoyant particle, let us consider the simple incompressible 
two-dimensional model flow defined by the stream function
\begin{equation}
\psi(x,y,t)=A \cos(x+B\sin\omega t)\cos y
.\label{flow}\end{equation}
The equations of motion for an element of the fluid will then be
$\dot{x}=\partial_y\psi$, $\dot{y}=-\partial_x\psi$.
$\psi$ has the r\^ole of a Hamiltonian for the dynamics of such an element,
with $x$ and $y$ playing the parts of the conjugate coordinate and momentum
pair. Let us first consider the simplest case where the time dependence is
suppressed, by setting $B=0$. Thence $\psi$ should be a constant of motion,
which implies that real fluid elements follow trajectories that are level
curves of $\psi$. Such level curves are depicted in Fig.~\ref{timeind}a, which
also shows contours of $Q$. The high values of $Q$ are around the
hyperbolic points, while negative $Q$ coincides with the centers of vortices
--- elliptic points --- in the flow. Figure~\ref{timeind}b shows the trajectory of a
neutrally buoyant particle starting from a point on a fluid trajectory within
the central vortex, with an inappreciable velocity mismatch with the flow. This
mismatch is amplified in the vicinity of the hyperbolic stagnation points where
$Q$ is larger than $4/9\,{\mathrm St}^{-2}$ to the extent that the particle 
leaves  the central vortex for one of its neighbors, a trip that is not allowed
to a  fluid parcel. In the end a particle settles on a trajectory, proper for a
fluid parcel, that does not visit regions of high $Q$. While this effect is
already seen in Fig.~\ref{timeind}b, it is more dramatically pictured in the
trajectory shown in Fig.~\ref{timeind}c, in which the particle performs a long
and complicated excursion wandering between different vortices before it
settles in a region of low $Q$ of one of them. To illustrate the divergence of
particle and fluid trajectories, and the fact that particle and fluid finally
arrive at an accord, in Fig.~\ref{timeind}d we display the difference between
the particle velocity and the fluid velocity at the site of the particle
against time for the case of Fig.~\ref{timeind}c. This difference is initially
inappreciable, and it converges to zero at long times, but during the 
interval in which the excursion takes place it fluctuates wildly. 

\begin{figure}
\begin{center}
\def\epsfsize#1#2{0.48\textwidth}
\leavevmode
\epsffile{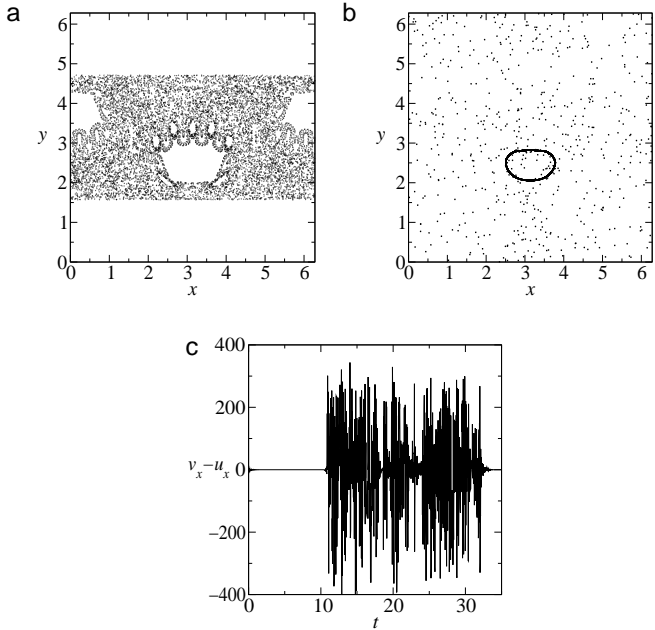}
\end{center}
\caption{\label{timedep}
Poincar\'e sections of trajectories in the time-dependent flow of 
Eq.\ (\ref{flow}) with $A=250$, $B=0.3$, and $\omega=1.0$.
(a) A chaotic fluid trajectory.
(b) The motion of a neutrally buoyant particle with Stokes number 
${\mathrm St}=0.2$ in the flow.
(c) The velocity difference $v_x - u_x$ between the particle and the flow 
against time.
}
\end{figure}

Even more interesting is the case of time-dependent flow: $B\neq 0$ in our
model. As in a typical Hamiltonian system, associated with the original
hyperbolic stagnation points, there are regions of the phase space, which is
here real space, dominated by chaotic trajectories. An individual trajectory of
this kind, stroboscopically sampled at the frequency of the flow, is reproduced
in Fig.~\ref{timedep}a. Such a trajectory visits a large region of the space,
which includes the original hyperbolic stagnation points and their vicinities
where $Q$ is large. Excluded from the reach of such a chaotic trajectory remain
areas where the dynamics is regular; the so-called KAM tori. In our model these
lie in the regions where $Q<4/9\,{\mathrm St}^{-2}$. Now a neutrally buoyant
particle that tries to follow a chaotic flow pathline will eventually reach the
highly hyperbolic regions of the flow. This makes likely its separation and
departure from such a pathline, in search of another pathline to which to
converge. However, convergence will only be achieved if the pathline never
crosses areas of high $Q$. Figures~\ref{timedep}b and \ref{timedep}c
demonstrate this phenomenon: a particle was released in the chaotic zone with
an inappreciable  velocity mismatch. The particle followed the flow, until,
coming upon a region of sufficiently high $Q$, it was thrown out of that flow
pathline onto a long excursion that finally ended up in a regular region of the
flow on a KAM torus. The regular regions of the flow then constitute
attractors of the dissipative dynamical system Eq.\ (\ref{disssys}) that
describes the behavior of a neutrally buoyant particle. The chaotic
trajectories in a Hamiltonian system are characterized by positive Lyapunov
exponents. The Lyapunov exponents are an average along the trajectory of the
local rate of convergence or divergence. Such a rate is measured by the
quantity $\lambda$. Hence, for a trajectory to be chaotic, it is a necessary
condition that it visit regions of positive $Q$: an upper bound to $Q$ is an
upper bound to the Lyapunov exponent.

\begin{figure}
\begin{center}
\def\epsfsize#1#2{0.48\textwidth}
\leavevmode
\epsffile{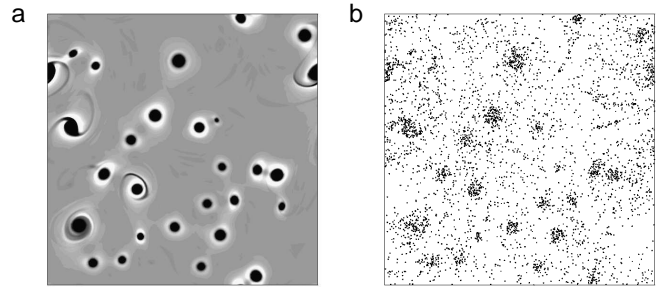}
\end{center}
\caption{\label{turbulent}
Small neutrally buoyant tracer particles with Stokes number 
${\mathrm St}=0.2$ collect in the centers 
of vortices in a two-dimensional turbulent flow simulation. 
(a) $Q$ field at time $t=1$ (lighter shading is higher $Q$), 
(b) distribution at time $t=1$ of particles that were uniformly
distributed in the flow at time $t=0$.
}
\end{figure}

We have considered the implications of these results for two-dimensional
turbulent flows, in which $Q$ defines three regions:
in the vortex centers it is strongly negative; in the circulation cells that 
surround them, strongly positive, while in the background between vortices 
it fluctuates close to zero \cite{elhmaidi,basdevant,hua,protas,provenzale}. 
We solve the two-dimensional vorticity equation for an incompressible fluid,
$\partial_t \omega + J(\omega,\psi) = D_\omega$, where $D_\omega=-\nu \nabla^4
\omega$ represents subgrid-scale dissipation; see, e.g., McWilliams
\cite{mcwilliams}. We integrate this equation on a doubly periodic domain using
a pseudospectral method with $512^2$ collocation points and $\nu=2.5\cdot
10^{-7}$; see Montabone \cite{montabone} and Provenzale \cite{provenzale} for
details. As a result of the dynamics, an initially uniform distribution of
small neutrally buoyant particles evolves in time towards an
asymptotic distribution concentrated in the inner part of vortices where $Q<0$,
and with voids in the areas crossed by fluid trajectories that visit regions
where $Q>4/9\,{\mathrm St}^{-2}$, as we illustrate in Fig.~\ref{turbulent}. 
This has importance consequences for the design of both fluid experiments using
neutrally buoyant tracer particles, and Lagrangian drifters in geophysical
flows.

Now let us consider the following problem in the realm of Hamiltonian
dynamics: given a generic chaotic Hamiltonian 
$H(p_1,\ldots p_n,q_1,\ldots q_n)$,
we might like to find orbits with Lyapunov exponents with a given bound. In
particular, we may be interested in locating small KAM tori in a chaotic sea.
Inspired by the above results, we may follow the dynamics of a small
neutrally buoyant hypersphere in a 2$N$-dimensional hyperfluid. By this, we
mean a particle that follows a simplified Maxey--Riley equation, 
Eq.\ (\ref{neutral}), extended to 2$N$ dimensions, which leads to a 
generalization of Eq.\ (\ref{disssys}),
$d{\mathbf A}_D/dt=M\cdot{\mathbf A}_D$,
with elements
$M_{ij}=\delta_{i,j}[(-1)^{i-1}\lambda_{[(i+1)/2]}-2/3\,{\mathrm St}^{-1}]$,
 where $[.]$ denotes the integer part, replacing Eq.\ (\ref{AD}). Thus the
Stokes number retains the same control over the Lyapunov exponents as in two
dimensions. The quantity $\mathbf A$ may be viewed as the control signal: it
vanishes when the desired trajectory is reached. In this light, the dynamics of
a neutrally buoyant particle as demonstrated above may be thought of as a tool
for targeting in Hamiltonian systems.

\begin{figure}
\begin{center}
\def\epsfsize#1#2{0.48\textwidth}
\leavevmode
\epsffile{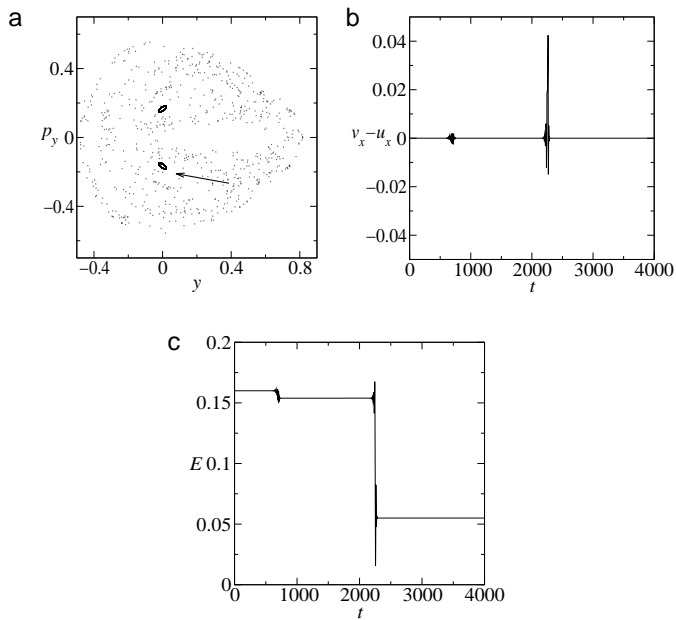}
\end{center}
\caption{\label{ham_control}
(a) A particle in a hyperfluid defined by the H\'enon--Heiles Hamiltonian
finds a KAM torus (arrowed) in the flow.
(b) The velocity difference $A_x=v_x - u_x$ between the particle and the flow 
against time.
(c) The H\'enon--Heiles energy $E$ against time.
}
\end{figure}

In Fig.~\ref{ham_control} we illustrate our targeting mechanism with the 
H\'enon--Heiles Hamiltonian $H=1/2(x^2+y^2+p_x^2+p_y^2)+x^2y-y^3/3$. Figure
\ref{ham_control}a shows with a Poincar\'e section how a particle released into 
a H\'enon--Heiles flow ends up on a KAM torus. In Fig.~\ref{ham_control}b we 
plot one component of the velocity difference between particle and flow; it
displays two episodes of particle--flow separation before the particle
settles on a KAM torus. Since it is not obliged to conserve the 
H\'enon--Heiles energy, the energy at which the particle finally settles into
the Hamiltonian flow is a priori undefined; Fig.~\ref{ham_control}c shows 
a series of plateaux punctuated by rapid energy jumps correlated with the
separations. If one is interested in finding a KAM torus at a given energy,
this constraint should be imposed on the dynamics of the particle.

We are grateful to Luca Montabone for
Fig.~\ref{turbulent}, and to Bill Young for useful discussions. JHEC and OP
acknowledge the financial support of the Spanish DGICyT, contract PB94-1167, 
and CICyT contract MAR98-0840. JHEC also thanks the Spanish CSIC for financial
support. This work was completed at the European Science Foundation TAO
(Transport processes in the Atmosphere and Oceans) Study Centre held in Palma
de Mallorca, Spain, during September 1999; we thank the ESF for its support.

\end{twocolumns}
\end{document}